\documentclass[12pt]{article}
\usepackage{cite}
\newlength{\dinwidth}                       
\newlength{\dinmargin}                      
\def\lsim{\mathrel{\rlap{\lower4pt\hbox{\hskip1pt$\sim$}}
    \raise1pt\hbox{$<$}}}                
\def\gsim{\mathrel{\rlap{\lower4pt\hbox{\hskip1pt$\sim$}}
    \raise1pt\hbox{$>$}}}                
\begin{document}

\renewcommand{\thefootnote}{\fnsymbol{footnote}}
\setcounter{footnote}{0}
\footnotetext[0]{\it Electronic mail addresses:\\
erdmann@dice2.desy.de,
Dirk.Graudenz@cern.ch,
leif@nordita.dk,
toku@vxdesy.desy.de
}
\renewcommand{\thefootnote}{\arabic{footnote}}
\setcounter{footnote}{0}

\vspace*{1cm}
\begin{center}  \begin{Large} \begin{bf}
Jets and High-{\boldmath$E_\perp$} Phenomena\\
  \end{bf}  \end{Large}
  \vspace*{5mm}
  \begin{large}
Martin Erdmann$^a$, Dirk Graudenz$^b$, 
Leif L{\"o}nnblad$^c$, Katsuo Tokushuku$^d$\\ 
  \end{large}
\end{center}
$^a$ Physikalisches Institut, 
     Universit\"at Heidelberg, 
     Philosophenweg 12, D-69120 Heidelberg, FRG\\
$^b$ CERN, Theory Divison, 
CH-1211 Gen{\`e}ve 23, Switzerland\\
$^c$ NORDITA, Blegdamsvej 17, DK-2100 Copenhagen \O, Denmark\\
$^d$ Institute for Nuclear Study, University of Tokyo, Tanashi, Tokyo 188,
     Japan\\
\begin{quotation}
\noindent
{\bf Abstract:} The working group on jets and high-$E_\perp$ phenomena
studied subjects ranging from next-to-leading order (NLO) corrections
in deeply inelastic scattering (DIS) and photoproduction with the
corresponding determinations of physical quantities, to the physics of
instanton-induced processes, where a novel non-perturbative
manifestation of QCD could be observed. Other centres of interest
were the physics of the forward direction, the tuning of event
generators and the development of a new generator which includes a
consistent treatment of the small- and large-$x$ QCD evolution.
The recommendations of the working group concerning detector upgrades
and machine luminosity are summarized.
\end{quotation}
\section*{Introduction}
The physics of hadronic final states is currently one of the main
interests at HERA. To mention only a few points, the study of jets has
led to a determination of 
the strong coupling constant 
and of the gluon density, and the
investigation of the hadronic activity in the forward direction has
improved our understanding of parton radiation in the initial
state. Concurrently with these phenomenology issues, there was the
development of tools such as next-to-leading order Monte Carlo
programs for jet production and event generators modelling the hadronic
final state.  The goal of the working group was to study the future
prospects of the physics of jets and high-$E_\perp$ phenomena in the
light of the two different improvements of an increased machine
luminosity (of the order of $\int{\cal
L}\mbox{d}t=250-1000\,\mbox{pb}^{-1}$) and improved detectors in the
forward direction.  Because of the wide range of subjects, the
working group was organized in four subgroups:
\begin{itemize}
  
\item{\bf\boldmath Deeply Inelastic Scattering.} The subjects
  considered in this subgroup were the study of QCD-instanton-induced
  processes, the calculation of jet cross-sections in NLO and the
  extraction of the strong coupling constant and the gluon density
  via hadronic final states.  A particular emphasis has been the study
  of the statistical and systematic errors for large luminosity.  One
  project studied the semi-DIS region, defined by events with
  $p_\perp \gg Q \gg \Lambda_{\mbox{\scriptsize QCD}}$, and the
  prospects of the determination of the virtual photon structure
  function.
  
\item{\bf QCD Evolution and the Forward Region.} This subgroup studied
  the prospects of measuring in deeply inelastic scattering the QCD
  evolution in the initial state.  Several small groups searched for
  relevant observables in order to (a) distinguish the QCD evolution
  schemes of DGLAP and BFKL, (b) detect instanton formation, and (c)
  establish `hot spots' in the proton.  Also studied were detector
  upgrades in the outgoing proton direction which concern the results
  of this working group and the working group on {\it Diffractive Hard
  Scattering}. All results from the two working groups which are
  related to a detector upgrade in the forward direction are
  summarized in a separate report \cite{upgrade}.
  
\item{\bf\boldmath Photoproduction.} Two projects considered the
  calculation of the NLO corrections to jet cross-sections, where in
  one of the projects the matching of theoretical and experimental jet
  cross-sections has been studied in detail. The measurement of the
  gluon density of the photon by means of the rapidity distribution of
  charged particles has been studied. Two projects considered the
  effects of colour coherence and of rapidity gaps between jets,
  respectively, and one project studied prompt photon, Drell--Yan and
  Bethe--Heitler processes.
  
\item{\bf\boldmath Event Generators and Tuning.} In this subgroup, a
  standardized framework ({\sc HZTool}) for the comparison of
  experimental data and generator predictions has been developed and
  used to tune existing generators. Another project considered the
  implementation of the linked dipole chain model in a Monte Carlo program
  interfaced to {\sc Ariadne}.

\end{itemize}

The outline of this working group summary is as follows.  The next
section introduces the notation.  In the following four sections the
activities of the subgroups are summarized.  A concluding section then
gives the final recommendations of the working group concerning
detector upgrades and machine luminosity.

\section*{Notation}

The momenta of the incident and outgoing electron\footnote{We use the
  term ``electron'' as synonymous to ``positron''. Charged-current
  processes and $Z^\circ$~exchange have not been studied in the
  working group.} and of the incident proton are denoted by $l$,
$l^\prime$ and $P$, respectively.  In deeply inelastic scattering, the
electron phase space is parametrized by the Bjorken variable\footnote{
  The variable $x_B$ is sometimes also denoted by $x$.  }
$x_B=Q^2/2Pq$ and by $y=Pq/Pl$, where $q=l-l^\prime$ is the
(space-like) momentum of the exchanged virtual photon, and $Q^2=-q^2$
is the square of the photon virtuality. In this way $Q$ represents the
energy scale of the scattering and $x_B$ may be interpreted, in the
case of lowest-order QCD sub-processes, as the momentum fraction of
the proton carried by the scattered parton.  For some processes such
as heavy-flavour production or the photoproduction of large transverse
energy jets the energy scale is not determined by the photon
virtuality.  In these cases the photon virtuality may be denoted by
$P$ (not to be confused with the proton momentum), where for
photoproduction, $P^2 \approx 0$.

Because of the hadronic component of a real photon, the parton
densities $f_{i/\gamma}(x_\gamma,\mu^2)$ of the photon have to be
introduced for photoproduction. Here $x_\gamma$ denotes the momentum
fraction of the photon carried by the parton~$i$.  More commonly used
is the experimentally observed quantity $x_\gamma^{\mbox{\tiny OBS}}
\equiv \left(E_{\perp 1} e^{-\eta_1} + E_{\perp 2}
  e^{-\eta_2}\right)/2yE_l$ derived from the two jets with the highest
$E_{\perp}$.

The convention of placing the proton along the positive $z$-axis in
the laboratory frame, and the virtual photon along the positive
$z$-axis in the hadronic centre of mass and Breit frames is used
throughout. For jet rates in DIS the ``(n+1)'' counting convention is
used, where the ``+1'' refers to the proton remnant jet.

References to contributions printed in these proceedings are made
by quoting the {\it author names}, printed in italics.

\section*{Deeply Inelastic Scattering}
The subgroup on deeply inelastic scattering had three main focuses:
QCD instantons, the calculation of next-to-leading-order jet
cross-sections, and the determination of the strong coupling constant
$\alpha_s(\mu_r^2)$ and the gluon density $g(x,\mu_f^2)$ from hadronic
final states.  Deeply inelastic scattering is defined by a photon
virtuality much larger than the fundamental QCD scale parameter,
$Q\gg\Lambda_{\mbox{\scriptsize QCD}}$. The presence of this large
scale allows the calculation of infrared-safe quantities in
perturbative QCD.  A possible approach to the inclusion of
hadronization effects in an analysis is to take them into account by
data unfolding or by the inclusion of correction factors based on a
comparison of the hadron level and the parton level by means of event
generators.

{\bf QCD Instantons.} QCD instantons give rise to helicity-violating
non-perturbative processes, whose experimental discovery would clearly
be of basic significance.  {\it M.~Gibbs, T.~Greenshaw, D.~Milstead,
  A.~Ringwald and F.~Schrempp} considered the discovery potential for
these processes at HERA by studying the characteristics of the
hadronic final state.  Because the processes are flavour-democratic,
strange particles would be produced in abundance. In addition, a
suitably defined event-shape variable might help to discriminate the
QCD-instanton-induced processes from standard QCD background.  Despite
large uncertainties in the first (preliminary) estimates of the
cross-section, HERA offers a distinct discovery window for these
spectacular processes, notably with a substantial luminosity upgrade.

{\bf NLO Corrections.} The calculation of jet cross-sections in NLO
was considered in two projects.  {\it E.~Mirkes and D.~Zeppenfeld}
have calculated the (2+1) jet cross-section by means of the phase
space slicing method, employing helicity amplitudes and the technique
of universal crossing functions. {\it S.~Catani and M.~Seymour} used
the subtraction method, where the subtraction term in the collinear
and soft regions is obtained by means of the recently developed dipole
formalism.  Because the Monte Carlo program based on the latter
calculation has been finished only recently, a numerical comparison of
the two different approaches has not yet been done.

{\bf The Strong Coupling Constant.} The future prospects of the
determination of $\alpha_s$ via the (2+1) jet rate has been considered
by {\it Th.~Hadig, Ch.~Niedzballa, K.~Rabbertz and K.~Rosenbauer}.
They studied the dependence of statistical and systematic errors in
dependence of the available luminosity. It turns out that the energy
scale error of the detector is the dominant experimental systematic
error. Assuming this error to be $2\%$, a total error of $\pm0.007$
can be achieved for $\alpha_s(M_Z^2)$ with $\int{\cal
  L}\mbox{d}t=250\,\mbox{pb}^{-1}$, which is to be compared with the
present error of the world average of $\alpha_s(M_Z^2)$ of $\pm0.006$.
A further increase of the luminosity might lead to a reduction of the
energy scale error and thus to a further improvement of the error. The
effect of additional acceptance cuts to reduce the systematic error
has also been studied.  In particular, a cut in the jet transverse
momentum seems to be promising. The systematic error induced by the
dependence of parton densities on $\Lambda_{\mbox{\scriptsize QCD}}$
has been estimated by {\it J.~Ch{\'y}la and J.~Rame\v{s}} by
considering the relative importance of $\alpha_s$ in the matrix
element and in the evolution of the parton densities. At moderately
large $Q^2$, where the present $\alpha_s$~measurements have been done,
the former is dominant. It would be desirable to find a way to
consistently include this dependence at smaller $Q^2$, where the data
sample is much larger.  An $\alpha_s$ measurement by means of scaling
violations of fragmentation functions has been studied by {\it
  D.~Graudenz}. Here a large systematic error is induced by the choice
of parton densities.  This error can be reduced by going to large
values of~$Q^2$.  Because of the rapid fall-off of the cross-sections,
a large luminosity is required. It turns out that the measurement
would not be competitive concerning the size of the error (the effect
being only logarithmic in the factorization scale), but might be an
interesting complementary measurement at HERA.

{\bf The Gluon Density.} The photon--gluon fusion process, giving rise
to (2+1) jets in the final state, can be exploited for a measurement
of the gluon density.  {\it G.~Lobo} has studied the prospects for a
combined global fit of $F_2$ and jet rates.  By including the jet rate
data, the error at large~$x\gsim 0.03$ can be reduced considerably.
The global approach also allows a combined fit of the quark and gluon
densities.  The direct measurement of $g(x)$ by means of the Mellin
transform method has been studied by {\it D.~Graudenz, M.~Hampel and
  A.~Vogt}. Here the quark densities are assumed to be input
distributions; the momentum sum rule is taken care of by means of the
normalization of $g(x)$. An increased luminosity of the order of
$250\,\mbox{pb}^{-1}$ may allow the reduction of the error band by a
factor of two, compared to the present integrated luminosity of
$3\,\mbox{pb}^{-1}$.

{\bf The Semi-DIS Region.} {\it J.~Ch{\'y}la and J.~Cvach} have studied
the prospects of a measurement of the virtual photon structure
function by looking at DIS events with some additional hard scale
$p_\perp \gg Q \gg \Lambda_{\mbox{\scriptsize QCD}}$, and conclude
that an integrated luminosity of $50\,\mbox{pb}^{-1}$ is sufficient
for a measurement that allows for a discrimination between various
models, assuming the virtual photon structure functions suppression is
$x$-independent. To measure the $x$-dependence of the virtual photon
structure functions, an integrated luminosity at least 10 times higher
would be necessary.

Except for the analysis in the semi-DIS region, all projects in this
subgroup related to the extraction of physical quantities as well as
the QCD instanton study strongly favour a substantial luminosity
increase, whereas a detector upgrade in the forward direction is not
required.  The $\alpha_s$~analysis shows that above an integrated
luminosity of about $250\,\mbox{pb}^{-1}$ the systematic errors will
eventually dominate over the statistical ones. A similar situation can
be expected in the case of the direct determination of the gluon
density via jets. It should be kept in mind, however, that the energy
scale error, and thus the systematic error of the extracted physical
quantities, also depends on the available integrated luminosity, since
high-$p_\perp$ jets are required to calibrate the detector
\cite{gayler96}.

\section*{QCD Evolution and the Forward Region}

The leading question of the `forward physics' group was: how can we
understand the QCD evolution of the initial state?  Compared to the
interpretation of inclusive measurements of the proton structure
function $F_2$, exclusive measurements in the forward direction
(outgoing proton) are sensitive to the explicit details of the
evolution between the proton and the photon--quark vertex.

Today's conventional description of the evolution of a single parton
are the DGLAP evolution equations.  These equations resum terms of the
form $(\alpha_s \ln Q^2)^n$.  At small fractional parton momenta~$x$,
contributions of the form $(\alpha_s \ln (1/x))^n$, not described by
the DGLAP equations, become important.  It is, however, still debated
at which values of~$x$ this will be the case.  HERA offers the
opportunity to settle this question empirically, for instance by
testing predictions of the BFKL type against those of DGLAP evolution.
Apart from these perturbatively calculated effects, also
non-perturbative effects, such as instanton formation, are expected to
contribute to the parton evolution.  Beyond single parton evolution,
multi-parton evolution effects are expected which could exhibit
inhomogeneous regions in the proton, e.g.\ regions of high parton
density (`hot spots'). Their detection would imply a significant step
forward in our understanding of the proton.

It is essential to find observables which reflect the evolution
of the partons from the proton to the $\gamma q$ vertex:
\begin{description}
\item[a)] \hspace{0.3cm} Indirect access is given by a measurement of
  kinematical variables of the final-state proton which can, in the
  case of a hard scattering process, be described by models where the
  proton initially lost partons during the scattering and finally
  received partons for the colour neutralization process ({\it
    I.~Gialas, J.~Hartmann; A.~Edin, G.~Ingelman and J.~Rathsman}).
\item[b)] \hspace{0.3cm} Direct measurements of the parton evolution
  require observables which involve high transverse momenta in order
  to suppress the influence of non-perturbative effects ({\it A.~Edin,
    G.~Ingelman and J.~Rathsman}).  Single charged particle spectra
  can distinguish at high transverse momenta different scenarios of
  QCD evolution ({\it M.~Kuhlen}).  In a similar way, jet
  cross-sections can be used to study parton evolution in the forward
  direction ({\it T.~Haas and M.~Riveline; J.~Bartels, A.~De~Roeck and
    M.~W\"usthoff}).  In a related project ({\it E.~Mirkes,
    D.~Zeppenfeld}) it has been found that the measured forward jet
  cross section at small $x_B$ \cite{H1fjet} is not described by a
  fixed-order NLO calculation. However, it can be explained by a LO
  calculation amended with a BFKL ladder in the initial state
  \cite{Bartels}. A different class of observables are shape variables
  which can, in principle, resolve short range effects at sufficiently
  large transverse momenta ({\it H.~He\ss ling}).
\end{description}

Detector upgrades in the direction of the outgoing proton will give
essential improvements in all the direct measurements of parton
evolution ({\it M.~Kuhlen; A.~Bamberger, S.~Eisenhardt, H.~He\ss ling,
  H.~Raach and S.~W\"olfle}).  The extension of the ZEUS detector by a
PLUG calorimeter which enlarges the rapidity coverage by $1.6$ units,
was studied in detail ({\it A.~Bamberger et al.}).

A high luminosity upgrade of the HERA machine, as proposed by the
working group {\it HERA Optics and Layout of Interaction Region}, will
signal the end of the physics described in this section.  Before such
upgrade, data corresponding to a luminosity of order $100\,$pb$^{-1}$
should be collected in order to ensure that the HERA project may
contribute significantly to the understanding of QCD evolution.

\section*{Photoproduction}

A further, important field of testing perturbative QCD is the study of
photoproduction processes with large transverse energy in the hadronic
final state.  Here one of the goals is to obtain new information on
the partonic structure of the photon and the proton. Whereas
$F_2^{\gamma}$ measurements in $e^+e^-$ collider experiments constrain
the quark distribution in the photon, the gluon distribution is
largely unknown.  In the region where $x_{\gamma}$ is close to zero or
unity, even the quark distribution is not well constrained at present.
In this workshop, three different final states were considered to study the
structure of the photon and proton: jets, inclusive hadronic particle
distributions, direct photons and lepton pairs.

{\bf Jets.} {\it J.~Butterworth, L.~Feld, M.~Klasen and G.~Kramer}
have made a detailed comparison in order to match the definition of
jets in experimental and theoretical studies.  It is shown that one
can match jets of NLO calculations to various experimental jet
definitions by tuning a parameter $R_{sep}$. The matching is better
when the $E_\perp$ of jets becomes larger.  Smearing effects from
hadronization are smaller at high $E_\perp$ as well.  By selecting
high-$E_\perp$ jets in a good detector acceptance region
($E_\perp^{jet} \geq 30\,\mbox{GeV}$ and $\eta^{jet} <2$), one can
test the photon and proton structure in the high-$x$ region provided a
large integrated luminosity ($\geq 250\,\mbox{pb}^{-1}$) is available.
{\it P.~Aurenche, L.~Bourhis, M.~Fontannaz and J.Ph.~Guillet} have
developed a Monte Carlo program describing the 2- and 3-jet
photoproduction in NLO.  These jet cross-sections can be extracted
from the generated events using a cone algorithm together with desired
experimental cuts.

{\bf Inclusive Particle Production.} {\it J.~Binnewies, M.~Erdmann,
  B.~Kniehl and G.~Kramer} have demonstrated that inclusive
differential rapidity cross-sections of charged particles with large
transverse momenta are sensitive to the gluon distribution of the
photon, at low $x_{\gamma}$. Assuming that the gluon fragmentation
function will be better known from LEP data on the longitudinal
polarized cross-sections, the extraction of the NLO gluon distribution
in the photon can be done with a precision of the order 10\% using an
integrated luminosity of $100\,\mbox{pb}^{-1}$.

{\bf Prompt Photon and Lepton Pair Production.} With a high integrated
luminosity, it is possible to study the quark distribution of the
photon using processes which are suppressed by the fine structure
constant $\alpha$ relative to the dominant di-jet production, but with
the advantage of a very clean environment.  Processes with
non-hadronic particles directly coming from the hard scattering
process is an example of this.  {\it P.~Bussey} estimated the event
rates with a high-$p_\perp$ photon (prompt photon). One can obtain
sufficient data to determine the quark distributions in the photon at
the 5--10\% level with an integrated luminosity of
$1000\,\mbox{pb}^{-1}$.  {\it B.~Levtchenko and A.~Shumilin} studied
Drell--Yan process. It is important to separate Drell--Yan lepton
pairs from the pairs that come from the Bethe--Heitler process, which
has a much larger cross-section.  Kinematical cuts for this separation
are proposed.

There are two studies on event topology of high-$p_\perp$ photoproduction,
on the colour coherence effect and on colour-singlet exchange.
 
{\bf Colour Coherence.} {\it L.~Sinclair and E.~Strickland} studied
the effect of colour coherence in multi-jet events.  In order to
obtain a large sample of multi-jet events it is necessary to have high
luminosity. However, it turned out that a large acceptance in the
forward region is also important.  The luminosity upgrade at the
expense of reducing the forward region acceptance is not worthwhile
for this study.  By extending the detector coverage up to 4 in
pseudorapidity, the effect could be more pronounced.  {\it A.~Lebedev
and J.~Vazdik} studied the process-dependence of colour coherence.
Particle flows in the inter-jet region are sensitive to the effect.

{\bf Colour-Singlet Exchange.} {\it J.~Butterworth, M.~Hayes,
  M.~Seymour and L.~Sinclair} studied events with a rapidity gap
between jets.  The colour-singlet exchange appears here at a scale
where perturbative QCD calculations give reliable predictions.  Such
data therefore give access to the origin of the so-called hard
Pomeron.  It is shown that with a larger detector coverage in the
forward region one can obtain an unambiguous signature for the 
colour-singlet exchange. The luminosity requirement is of the order of
$100\,\mbox{pb}^{-1}$.

\section*{Event Generators and Tuning}

The importance of theoretically well-founded event generators which
give a good description of data cannot be emphasised enough. The
situation at HERA in this respect has not been very satisfactory,
especially when comparing to the extraordinary success of event
generators at LEP \cite{LEPtune}. This can be exemplified in DIS where
all available generators have had great problems with describing
fairly simple distributions, such as the $E_\perp$-flow \cite{Etflow}.

This is not surprising considering the extra complications introduced
at HERA which are not present in $e^+e^-$ annihilation. In
photoproduction there is the problem that the photon sometimes behaves
as a point-like object and sometimes as a resolvable hadron. In the
latter case, multiple interactions may occur, giving rise to an
underlying event. Also, in photoproduction as well as in DIS, there is
the problem of initial-state QCD evolution and how to handle it,
especially in the small-$x$ region, and in relation with the
fragmentation of the proton remnant.

During the course of this workshop, the situation has much improved.
Both for DIS and photoproduction, the available generators have been
developed and the agreement with data is now at a level where a tuning
of the parameters is meaningful. Two closely related projects have
been working with the tuning of event generators. One of them {\it
  (J.~Bromley, N.~Brook, A.~Buniatian, T.~Carli, G.~Grindhammer,
  M.~Hayes, M.~Kuhlen, L.~L\"onnblad and R.~Mohr)} developed a library
of FORTRAN routines called {\sc HZTool} for easy comparison of event
generators with published data. The other {\it (N.~Brook, T.~Carli,
  R.~Mohr, M.~Sutton and R.G.~Waugh)} used this library to perform a
first tuning of the DIS generators {\sc Ariadne}\cite{Ariadne}, {\sc
  HERWIG}\cite{HERWIG} and {\sc Lepto}\cite{Lepto}. A few of the
measured distribution were selected, from which a global $\chi^2$ was
constructed to measure the quality of the fits. For all three programs
the $\chi^2$ was much improved by the tuning. The final numbers
presented were $\chi^2 =$ 0.81, 1.85 and 1.36 per degree of freedom
for {\sc Ariadne}, {\sc HERWIG} and {\sc Lepto}, respectively.

One new generator has been developed by {\it G.~Gustafson,
  H.~Kharraziha and L.~L\"onnblad}. It implements the Linked Dipole
Chain model, which is a reformulation on the CCFM evolution equations
based on the colour dipole picture. Here a careful division between
initial and final state emissions results in a model well suited for
an event generator implementation. This is the first complete
generator where QCD coherence is correctly taken into account in the
small-$x$ region and where DGLAP and BFKL dynamics both are reproduced
in the relevant limits. The preliminary comparison with data presented
here looks promising.

\section*{Summary of Recommendations}

The recommendations of the working group concerning a detector upgrade
in the forward direction and an upgrade of the luminosity, as
summarized in Table~\ref{thetable}, are not unambiguous, because the
two options are conflicting.  The physics involving processes with
large transverse momenta of final state jets would benefit
substantially from a luminosity upgrade.  This would allow the study
of processes at large $Q^2$, thus increasing, for instance, the lever
arm for a measurement of the {\it running} coupling constant
$\alpha_s(Q^2)$.  A larger data sample would also permit the
application of strict acceptance cuts to bias the sample towards a
phase region where perturbative QCD is applicable unambiguously, i.e.\ 
without taking into account large hadronization or resummation
effects.  It would moreover allow a reduction of the energy scale
error, which has a direct impact on the extraction of physical
quantities.

On the other hand, HERA offers a unique opportunity to study the QCD
evolution and the physics of the forward direction in a comparably
clean environment.  The results of the working group show that most
studies in this phase space region would already benefit from an
increase of the detector acceptance by one unit of rapidity, with a
total integrated luminosity requirement of the order of
$100\,\mbox{pb}^{-1}$. However, the proposed high luminosity upgrade
of the HERA machine would to a large extent make such studies
impossible. In a separate report by members from both the {\it Jets
and High-$E_\perp$ Phenomena} and {\it Diffractive Hard Scattering}
working groups, the cases for a forward detector upgrade are
summarized \cite{upgrade}, strongly recommending that a luminosity
upgrade should at least be postponed to allow for more studies of
forward physics.

In conclusion: the physics of jets and high-$E_\perp$ phenomena will
continue to be a very interesting topic at HERA. Both options for the
future of HERA, a substantial luminosity increase and a forward
detector upgrade, which have been studied in the working group, would
mean new physics opportunities. It is worth while to consider running
HERA for two or three years with a total integrated luminosity of
$100\,\mbox{pb}^{-1}$, to allow for the instrumentation of and
measurement in the forward direction, and then moving to the
luminosity upgrade, which is definitely required for precise QCD
studies at large transverse momenta.

\begin{table}
\renewcommand{\baselinestretch}{1.5}
\begin{center}
\begin{small}
\begin{tabular}[h]{|c||l||c|c|c|c|}
\hline
\multicolumn{2}{|c||}{ }
 &
  &
   &
    & \\
\multicolumn{2}{|c||}{Jets and High-$E_\perp$}
 & \raisebox{-3.5ex}[-1.5ex]{Competition$^{(3)}$}
  & \hspace*{0.5cm}{\protect{$\int{\cal L}\mbox{d}t$}}\hspace*{0.5cm}  
   & Luminosity 
    & Detector \\
\multicolumn{2}{|c||}{Measurements$^{(1,2)}$}
 &  
  & $[{\rm pb}^{-1}]^{(4)}$ 
   & Upgrade$^{(5)}$    
    & Upgrade$^{(6)}$    \\
\multicolumn{2}{|c||}{ }
 &             
  & 
   &      
    &          \\ 
\hline\hline
 & Instanton ($K, \mu, E_\perp$)
  & 
   & 1000
    & yes
     &   \\  \cline{2-6}
 & $\alpha_s$ (particles, jets)
  & $e^+e^-$
   & 250
    & --
     &   \\  \cline{2-6}
 & gluon in proton (jets)
  & $\overline{p}p$
   & 250
    & --
     &   \\  \cline{2-6}
 & QCD evolution (jets) 
  & 
   & 30
    & no
     & ZEUS plug  \\ \cline{3-6}
\raisebox{1.5ex}[-1.5ex]{$DIS$}
 & 
  & 
   &
    &
     & very forward \\ 
 & \raisebox{2.0ex}[-1.5ex]{\hspace{1.2cm} (particles)}
  & 
   & \raisebox{2.0ex}[-1.5ex]{100}
    & \raisebox{2.0ex}[-1.5ex]{no}
     & H1 tracker \\  \cline{3-6}
 & \hspace{1.2cm} (shape variables) 
  & 
   & $>$10
    & no
     & ZEUS plug  \\  \cline{2-6}
 & proton fragm. (LPS)
  & 
   & $>$10
    & --
     &              \\ 
\hline\hline
$\gamma p\leftrightarrow$
 & quark in $\gamma^*$ (2 jets)
  & quark at LEPII
   & 
    & 
     &  \\ \cline{3-3}
$DIS$
 & gluon in $\gamma^*$ (2 jets)
  & none
   & \raisebox{2.0ex}[-1.5ex]{50/500}
    & \raisebox{2.0ex}[-1.5ex]{no}
     & \raisebox{2.0ex}[-1.5ex]{H1 VLQ}           \\ 
\hline\hline
 & gluon in $\gamma$ ($d\sigma^{\rm particle}/d\eta$)  
  & (charm at LEPII)  
   & 100      
    &  --
     &           \\ \cline{3-6}
 & quark in $\gamma$  (prompt $\gamma$) 
  &  
   &  1000 
    &  yes
     &           \\ \cline{4-6}
 &  \hspace{1.8cm}  (Drell--Yan) 
  &  quark at LEPII
   &  1000 
    &  yes
     &           \\ \cline{4-6}
$\gamma p$
 & \hspace{1.8cm}  (high-$E_\perp$ jets) 
  & 
   & 250
    & --
     &            \\ \cline{2-6}
 & hard Pomeron ($\Delta\eta_{\rm gap}^{\rm jets}$)
  & ($\overline{p}p$) 
   & 100      
    & no 
     & ZEUS plug \\ \cline{2-6}
 & colour coherence (jets)
  & $e^+e^-$, $\overline{p}p$
   & 250
    & no
     & ZEUS plug  \\ 
\hline
\end{tabular}
\end{small}
\end{center}
\renewcommand{\baselinestretch}{1}
\caption [recommendation] {\label{recommendation}
\label{thetable} 
Future HERA measurements recommended by the {\it Jets and
  High-$E_\perp$ Phenomena} working group;
the columns show \\
1) process type: deeply inelastic scattering (DIS),
photoproduction ($\gamma p$),\\
2) physics topic (experimental method);
LPS stands for the leading proton spectrometer, \\
3) (possible) competition from other laboratories,
\\
4) required luminosity in pb$^{-1}$, \\
5) statement on the luminosity upgrade {\em including
  acceptance losses in the current detectors}, \\
6) a detector upgrade which would significantly improve the results.
(VLQ = very low $Q^2$ tagger)}
\end{table}

\end{document}